\documentclass[%
reprint,
superscriptaddress,
amsmath,amssymb,
prc,
floatfix,]%
{revtex4-1}

\usepackage{CJK}
\usepackage{graphicx}
\usepackage{dcolumn}
\usepackage{bm}
\usepackage{enumerate}
\usepackage[bookmarks=true,colorlinks=true,linkcolor=blue,
           citecolor=blue,hypertex,breaklinks=true,urlcolor=blue]{hyperref}
\usepackage{booktabs}
\usepackage{array}
\begin{document}

\title{Ground-state properties of superheavy $Z=122$ isotopes within the
        deformed relativistic Hartree-Bogoliubov theory in continuum}

\begin{CJK*}{GBK}{}

 \author{Jin-Hong Zhuang }
 \affiliation{Mathematics and Physics Department,
              North China Electric Power University, Beijing 102206, China}

 \author{Zhen-Hua Zhang}
 \email{zhzhang@ncepu.edu.cn}
 \affiliation{Mathematics and Physics Department,
              North China Electric Power University, Beijing 102206, China}
 \affiliation{Hebei Key Laboratory of Physics and Energy Technology,
              North China Electric Power University, Baoding 071000, China}

 \author{Yuan-Yuan Wang }
 \affiliation{Mathematics and Physics Department,
              North China Electric Power University, Beijing 102206, China}
 \affiliation{Hebei Key Laboratory of Physics and Energy Technology,
              North China Electric Power University, Baoding 071000, China}

  \author{Cong Pan }
  \affiliation{Department of Physics, Anhui Normal University, Wuhu 241000, China}

  \author{Kai-Yuan Zhang }
  \affiliation{Institute of Nuclear Physics and Chemistry, China Academy of Engineering Physics,
               Mianyang, Sichuan 621900, China}

 \author{Huan-Yu Zhang }
 \affiliation{Mathematics and Physics Department,
              North China Electric Power University, Beijing 102206, China}

  \author{Yu Sun}
 \affiliation{Mathematics and Physics Department,
              North China Electric Power University, Beijing 102206, China}

\date{\today}

\begin{abstract}
The ground-state properties of superheavy $Z = 122$ isotopes are investigated using
the deformed relativistic Hartree-Bogoliubov theory in continuum (DRHBc).
Bulk properties, including binding energies,
Fermi energies, nucleon separation energies, two-neutron shell gaps,
quadrupole deformations, root-mean-square radii, and avarage pairing gaps are calculated.
The results are compared with those obtained from the
relativistic continuum Hartree-Bogoliubov (RCHB) theory.
By examining the dependence on the angular-momentum
cutoff and the effects of triaxial and octupole deformations,
a strategy for determining the ground states is suggested.
Furthermore, based on an analysis of the Fermi and nucleon separation energies,
the proton and neutron drip lines for $Z = 122$ isotopes
are determined within both the DRHBc and RCHB frameworks.
The possible magic numbers $N=184$, 258, and 350 are also suggested.
Finally, the evolution of the two-neutron shell gaps, deformation, charge and neutron radii,
single-particle levels, and average pairing gaps with increasing neutron number is discussed.
These quantities consistently support the suggested neutron shell closure.
\end{abstract}

\maketitle
\end{CJK*}

\section{Introduction}\label{sec:intro}

The synthesis of new superheavy elements is one of the most important research
frontiers in nuclear physics, as it helps us to address fundamental questions
regarding the boundaries of the nuclear chart and the limits of charge within atomic nuclei.
Experimentally, significant progress has been made in the synthesis of superheavy elements.
To date, those with atomic numbers $Z \le 118$ have been synthesized
through cold and hot fusion reactions~\cite{Hofmann2000_RMP72-733, Morita2004_JPSJ73-2593,
Oganessian2007_JPG34-R165, Oganessian2010_PRL104-142502,
Zhou2014_Wuli43-817, Zhou2017_NPR318-331, Jiang2025_Physics54-599, Huang2026_CPL43-010101}.
Currently, the synthesis of even heavier new superheavy elements,
specifically in the range of $Z = 119$-122,
represents a highly competitive frontier in nuclear research.
Most of these experiments are concentrating on $Z = 119$ and $120$.
Despite extensive attempts, no experiment has yet succeeded in synthesizing these
new elements~\cite{Oganessian2009_PRC79-024603,
Hofmann2016_EPJA52-116, Hofmann2016_EPJA52-180,
Khuyagbaatar2020_PRC102-064602, Sakai2022_EPJA58-238, Tanaka2022_JPSJ91-084201,
Gan2022_EPJA58-158, Gates2024_PRL133-172502, Zhang2025_NST36-204}.

Understanding the structure of the superheavy elements is quite helpful for synthesizing them.
Theoretically, extensive investigations have been conducted using various models,
such as macroscopic-microscopic models~\cite{Sobiczewski1966_PL22-500,
Meldner1967_ArkivF36-593, Nilsson1968_NPA115-545, Nilsson1969_NPA131-1,
Mosel1969_ZPA222-261, Moller1994_JPG20-1681, Mo2014_PRC90-024320, Wang2014_PLB734-215}
and self-consistent mean-field models~\cite{Bender2003_RMP75-121} including
non-relativistic~\cite{Cwiok1994_NPA573-356, Cwiok1996_NPA611-211,
Cwiok1999_PRL83-1108, Cwiok2005_Nature433-705, Robledo2018_JPG46-013001},
and relativistic~\cite{Lalazissis1996_NPA608-202, Zhang2005_NPA753-106,
Afanasjev2003_PRC67-024309, Li2014_PLB732-169, Agbemava2015_PRC92-054310,
Afanasjev2018_PLB782-533, Agbemava2019_PRC99-034316, Agbemava2021_PRC103-034323,
Taninah2020_PRC102-054330} density functional theories.
As for the superheavy region, there is little experimental information.
Therefore, the extrapolation power of these models is very important.

The deformed relativistic Hartree-Bogoliubov theory in
continuum (DRHBc)~\cite{Zhou2010_PRC82-011301R, Li2012_CPL29-042101, Li2012_PRC85-024312,
Zhang2020_PRC102-024314, Pan2022_PRC106-014316},
which can concurrently take into account pairing correlations,
continuum effects, and degrees of freedom in deformation, is one of the most potent models.
It can be viewed as the deformed counterpart of the relativistic continuum Hartree-Bogoliubov (RCHB)
theory~\cite{Meng1998_PRL80-460, Meng1996_PRL77-3963, Meng1998_NPA635-3, Meng2006_PPNP57-470},
and it has demonstrated remarkable ability in the satisfactory description of
ground-state properties through powerful explorations.
Based on the DRHBc, numerous interesting nuclear phenomena have been investigated,
such as the halo phenomena~\cite{Zhou2010_PRC82-011301R, Meng2015_JPG42-093101,
Sun2018_PLB785-530, Sun2020_NPA1003-122011, Sun2021_PRC103-054315,
Sun2021_SciBulletin66-2072, Zhong2022_SciChinaPMA65-262011, Zhang2023_PLB844-138112,
Wang2024_EPJA60-251, Zhang2024_PRC110-014320, Pan2024_PLB855-138792,
An2024_PLB849-138422, Zhang2025_PLB871-139989, Papakonstantinou2025_PRC112-044301},
the shape evolution and shape coexistence~\cite{Kim2022_PRC105-034340, Choi2022_PRC105-024306,
Guo2023_PRC108-014319, Zhang2023_PRC108-024310,
Mun2024_PRC110-024310, Mun2025_Particles8-32},
the shell evolution~\cite{Zhang2023_PRC107-L041303, Zheng2024_ChinPhysC48-014107,
Zhang2024_PRC110-024302, Huang2025_PRC111-034314},
fission barriers~\cite{Zhang2024_ChinPhysC48-104105},
$\alpha$-decay half-lives~\cite{Choi2024_PRC109-054310, Mun2025_Particles8-42},
the one-proton emission~\cite{Xiao2023_PLB845-138160, Lu2024_PLB856-138922}, etc.
In addition, the ground-state properties of those nuclei in the whole nuclear chart
have been calculated and the corresponding nuclear mass tables based on the
RCHB and DRHBc theories have been established~\cite{Xia2018_ADNDT121-122-1,
Zhang2022_ADNDT144-101488, Guo2024_ADNDT158-101661}.

In Ref.~\cite{Guo2024_ADNDT158-101661}, the ground-state properties of even-$Z$ nuclei
have been calculated systematically up to $Z=120$, which is predicted to be a proton magic number.
As discussed in Ref.~\cite{Indelicato2013_Nature498-40}, the upper limit of the nuclear charge
number may extend to $Z=173$ according to quantum electrodynamics calculations.
Therefore, it is essential to extend theoretical investigations beyond $Z=120$ within the DRHBc framework.
The DRHBc Mass Table Collaboration is now working on the whole nuclear chart
including the superheavy nuclei with $121 \le Z \le 136$.
In this work, we present the calculations for the ground-state properties across the entire $Z=122$
isotopic chain using the DRHBc theory, including binding energies, Fermi energies,
nucleon separation energies, quadrupole deformations, and root-mean-square (rms) radii, etc.
Note that IUPAC has already provided a temporary name ``Unbibium''
and symbol ``Ubb'' for the element with $Z=122$~\cite{Chatt1979_PAC5138-1384}.

This paper is organized as follows.
Section~\ref{sec:theor} provides a brief introduction to the framework of DRHBc.
The numerical details of the calculations are described in Sec.~\ref{sec:details}.
In Sec.~\ref{sec:result}, we present the ground-state properties of nuclei
in the $Z=122$ isotopic chain obtained from the DRHBc calculations,
with corresponding RCHB results included for comparison.
Finally, a brief summary is given in Sec.~\ref{sec:summary}.

\section{Theoretical framework}\label{sec:theor}

The details of the DRHBc can be found in Refs.~\cite{Li2012_PRC85-024312,
Chen2012_PRC85-067301, Zhang2020_PRC102-024314, Pan2022_PRC106-014316}.
Here we briefly present its formalism.
The relativistic Hartree-Bogoliubov (RHB) equation reads~\cite{Kucharek1991_ZPA339-23}
\begin{eqnarray}
\left(\begin{array}{cc}
	h_D-\lambda_\tau & \Delta \\
	-\Delta^* & -h_D^*+\lambda_\tau
\end{array}\right)\left(\begin{array}{l}
	U_k \\
	V_k
\end{array}\right)=E_k\left(\begin{array}{c}
	U_k \\
	V_k
\end{array}\right),
 \end{eqnarray}
where $\lambda_\tau$ is the Fermi energy for neutrons or protons ($\tau = n$ or $p$),
and $E_k$ and ($U_k$, $V_k$)$^T$ are the quasiparticle energy and wave function, respectively.
$h_D$ is the Dirac Hamiltonian,
\begin{eqnarray}
	h_D(\boldsymbol{r})=\boldsymbol{\alpha} \cdot \boldsymbol{p}+V(\boldsymbol{r})+\beta[M+S(\boldsymbol{r})],
\end{eqnarray}
where $M$ is the nucleon mass,
$S(\boldsymbol{r})$ and $V(\boldsymbol{r})$ are the scalar and vector potentials, respectively.
$\Delta$ is the pairing potential,
\begin{eqnarray}
	\Delta\left(\boldsymbol{r}_1, \boldsymbol{r}_2\right)=V^{\rm pp}\left(\boldsymbol{r}_1, \boldsymbol{r}_2\right)
\kappa\left(\boldsymbol{r}_1, \boldsymbol{r}_2\right),
\end{eqnarray}
where $\kappa$ is the pairing tensor~\cite{Ring2004_Book} and
$V^{\rm pp}$ is the pairing interaction in the particle-particle channel.
Here a density-dependent zero-range pairing force is adopted,
\begin{eqnarray}
V^{\rm pp}\left(\boldsymbol{r}_1, \boldsymbol{r}_2\right)=V_0 \frac{1}{2}\left(1-P^\sigma\right)
\delta\left(\boldsymbol{r}_1-\boldsymbol{r}_2\right)\left(1-\frac{\rho\left(\boldsymbol{r}_1\right)}{\rho_{\rm sat}}\right),
\label{Pairing}
\end{eqnarray}
in which $V_0$ is the pairing strength, $\rho_{\rm sat}=0.152$~fm$^{-3}$ denotes the saturation density of nuclear matter,
and $\frac{1}{2}\left(1-P^\sigma\right)$ represents the projector for the spin $S = 0$ component in the pairing channel.

In the DRHBc, since axial and reflection symmetries are assumed,
the pairing tensor, various densities and potentials are expanded in terms of the Legendre polynomials,
\begin{eqnarray}
f(\boldsymbol{r})=\sum\limits_{\lambda} f_\lambda(r) P_\lambda(\cos \theta), \quad \lambda=0,2,4,\ldots
\label{Legendre}
\end{eqnarray}

The RHB equations are solved in the Dirac Woods-Saxon (DWS)
basis~\cite{Zhou2003_PRC68-034323, Zhang2022_PRC106-024302},
which can provide equivalent description to coordinate-space solutions and
appropriately describe the large spatial extension of weakly bound nuclei.

For an odd-$A$ nucleus, one needs to further take into consideration the blocking effect
for the unpaired single proton or neutron~\cite{Li2012_CPL29-042101}.
The equal filling approximation is adopted to deal with
the blocking effects in the DRHBc~\cite{Pan2022_PRC106-014316}.

\section{Numerical details}\label{sec:details}

In this work, the relativistic density functional PC-PK1~\cite{Zhao2010_PRC82-054319} is adopted.
The DWS basis is constructed in a box of  $R_{\rm{box}} = 20$~fm, with a mesh of $\Delta r = 0.1$~fm.
For the basis space, the angular momentum cutoff is chosen as $J_{\rm max}=31/2~\hbar$,
and the energy cutoff in the Fermi sea is $E_\mathrm{cut}=300$~MeV.
The maximum expansion order in Eq.~(\ref{Legendre}) is $\lambda_{\rm max}=12$,
which is sufficient for our study.
For the particle-particle channel, the pairing strength $V_0=-300$~MeV$\cdot$fm$^{3}$
in Eq.~(\ref{Pairing}), and the pairing window is chosen as~100 MeV.
The examinations for the above numerical cutoffs and the pairing parameters
have been carried out for $Z=134$ and 135 isotopes~\cite{Wang2024_Particles7-1139}.
For comparison, the RCHB calculations are also performed for all the $Z=122$ isotopes considered.
The corresponding numerical details can be found in Ref.~\cite{Xia2018_ADNDT121-122-1}.

To examine the effects of octupole and triaxial degrees of freedom on the
potential energy curves (PECs), the calculations with multi-dimensionally constrained covariant
density functional theory (MDC-CDFT) are also carried out for $^{384}$Ubb as an example.
For comprehensive details on the MDC-CDFT framework, see Refs.~\cite{Lu2012_PRC85-011301R,
Lu2014_PRC89-014323, Zhou2016_PS91-063008, Zhao2017_PRC95-014320}.
In these calculations, the density functional PC-PK1 is also adopted.
The Dirac equation is solved by an expansion in the axially deformed harmonic
oscillator basis with $N_f=20$ major shells, which is enough for the superheavy nuclei.
Pairing correlations are taken into account in the BCS approximation with a finite range separable
pairing force~\cite{Tian2009_PLB676-44}.
The pairing strength and effective range are taken as $G/G_0 = 1.1$ and $a=0.644$~fm,
with $G_0 = 728$~MeV$\cdot$fm$^{3}$.
The pairing strength is obtained by reproducing the odd-even differences in binding energies
of the actinides and superheavy nuclei~\cite{Zhao2015_PRC91-014321, Meng2020_SciChinaPMA63-212011,
Wang2022_ChinPhysC46-024107, Deng2023_IJMPE32-2340004}.
Both in the DRHBc and MDC-CDFT, the microscopic center-of-mass correction
is considered~\cite{Bender2000_EPJA7-467, Long2004_PRC69-034319, Zhao2009-CPL26-112102}.

\section{Results and discussion}\label{sec:result}

The nuclei in the $Z=122$ isotopic chain with neutron number from $N=171$ to 352
are calculated by the DRHBc.
Bulk properties for the ground states, including binding energies, Fermi energies,
separation energies, quadrupole deformations, and rms radii, are obtained.
The results are also compared with those obtained by the spherical RCHB calculations.

\begin{figure*}[!]
\centering
\includegraphics[width=0.7\textwidth]{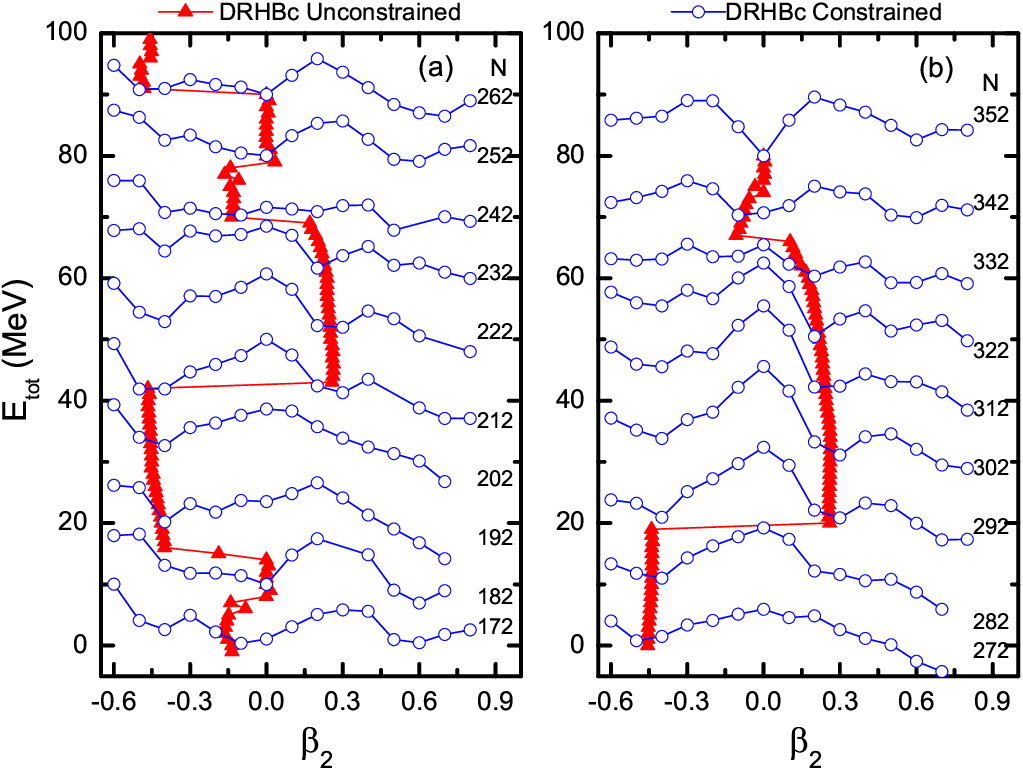}
\caption{The evolution of PECs for the $Z=122$ isotopic chain as a function of neutron number,
from $N=172$ to $N=352$, obtained from constrained DRHBc calculations using the density functional PC-PK1.
The calculations are performed at intervals of $\Delta N=10$.
In each panel, the PECs for $^{294}$122 ($N=172$) and $^{394}$122 ($N=272$) are renormalized
to the energy of their respective ground states (indicated by red solid triangles).
The remaining PECs are successively shifted upward by 1 MeV for each additional neutron.
The ground-state deformations determined from unconstrained DRHBc calculations
are marked by solid red triangles on the corresponding curves.
}
\label{fig:pec}
\end{figure*}

Figure~\ref{fig:pec} presents the evolution of the PECs
for the $Z=122$ isotopic chain, calculated from $N=172$ to $N=352$ using
constrained DRHBc calculations at intervals of $\Delta N=10$.
The ground-state deformations from unconstrained calculations are indicated by solid red triangles.
These ground states correspond to the global minima on their respective PECs,
validating the self-consistence of the DRHBc calculations.
It can be seen that, each PEC typically exhibits several local minima.
According to Ref.~\cite{Wang2024_Particles7-1139}, an angular momentum cutoff
of $J_{\rm max}=31/2~\hbar$ in the DRHBc has been verified
as sufficient for nuclei with proton numbers $Z=122$-136.
That study also recommended that, with $J_{\rm max}=31/2~\hbar$,
only the lowest minimum in the small-deformation region $|\beta_2| < 0.3$ can be reliably identified as the ground state,
whereas the lowest minimum at larger deformations $|\beta_2| > 0.3$ should be checked carefully,
because the PECs in the large-deformation region vary significantly as $J_{\text{max}}$ increases.
In the present work, numerous minima appear at large oblate ($\beta_{2} \sim -0.4$ to $-0.5$)
and prolate ($\beta_{2} > 0.5$) deformations.
It is therefore essential to determine which minimum in each PEC represents the actual ground state.

\begin{figure}[h]
\centering
\includegraphics[width=0.95\columnwidth]{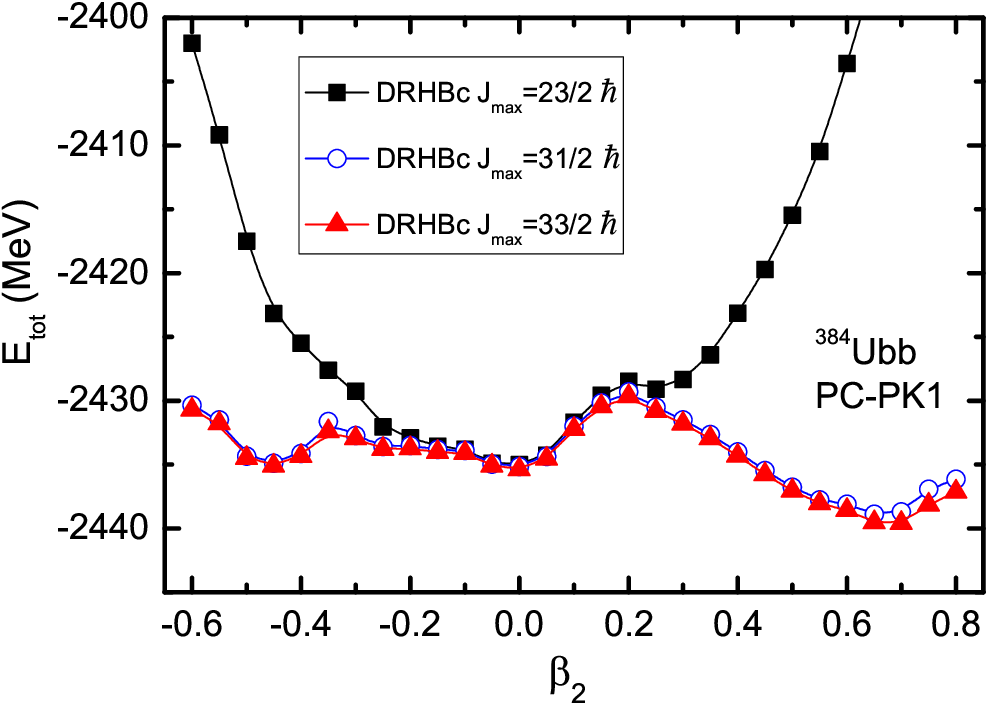}
\caption{The PECs of $^{384}$Ubb obtained by the DRHBc calculations
with the angular momentum cutoff $J_{\rm max}= 23/2~\hbar$ (black solid squares),
31/2~$\hbar$ (blue open circles) and 33/2~$\hbar$ (red solid triangles).
}
\label{fig:jm}
\end{figure}

The nucleus $^{384}$Ubb, which exhibits several minima in its PEC, is taken as an example
to show the strategy for determining the ground state.
Figure~\ref{fig:jm} shows the PECs of $^{384}$Ubb
calculated using the DRHBc with different angular momentum cutoffs:
$J_{\text{max}} = 23/2~\hbar$ (black solid squares), $31/2~\hbar$ (blue open circles),
and $33/2~\hbar$ (red solid triangles).
Note that for $J_{\text{max}} = 23/2~\hbar$, the pairing interaction strength
is set to $V_0 = -325$~MeV$\cdot$fm$^{3}$~\cite{Zhang2020_PRC102-024314}.
It can be seen that at $J_{\text{max}} = 23/2~\hbar$, only two minima are present:
one spherical minimum and a very shallow prolate minimum around $\beta_2 \sim 0.25$.
When $J_{\text{max}}$ increases to $31/2~\hbar$, the PEC drops significantly
in both the prolate and oblate large-deformation regions.
An obvious oblate minimum appears near $\beta_2 \sim -0.45$,
while the prolate minimum shifts to $\beta_2 \sim 0.65$.
This behavior is consistent with the findings in Ref.~\cite{Wang2024_Particles7-1139}
regarding the convergence with respect to $J_{\text{max}}$.
In addition, a very flat oblate minimum also appears near $\beta_2 \sim -0.25$.
Further increasing $J_{\text{max}}$ to $33/2~\hbar$ leads to nearly converged
PECs across most deformation regions,
except for the very large prolate deformation ($\beta_2 > 0.6$) shifted a little (about 0.05).
This indicates that an angular momentum cutoff of $J_{\text{max}} = 31/2~\hbar$
is sufficient for the present calculations.
Therefore, from the perspective of the axially symmetric and reflection symmetric DRHBc model,
the minimum with large oblate deformation $\beta_2 \sim -0.45$ should be regarded as the ground state,
given its lower energy and stability upon further increasing $J_{\text{max}}$.
This finding supersedes the previous identification of the minimum within $|\beta_2| < 0.3$
as the ground state~\cite{Wang2024_Particles7-1139}.
In contrast, whether the lowest minimum with large prolate deformation ($\beta_2 > 0.5$) corresponds
to the ground state remains uncertain since it changes slightly with further increasing $J_{\text{max}}$.

\begin{figure}[h]
\centering
\includegraphics[width=0.95\columnwidth]{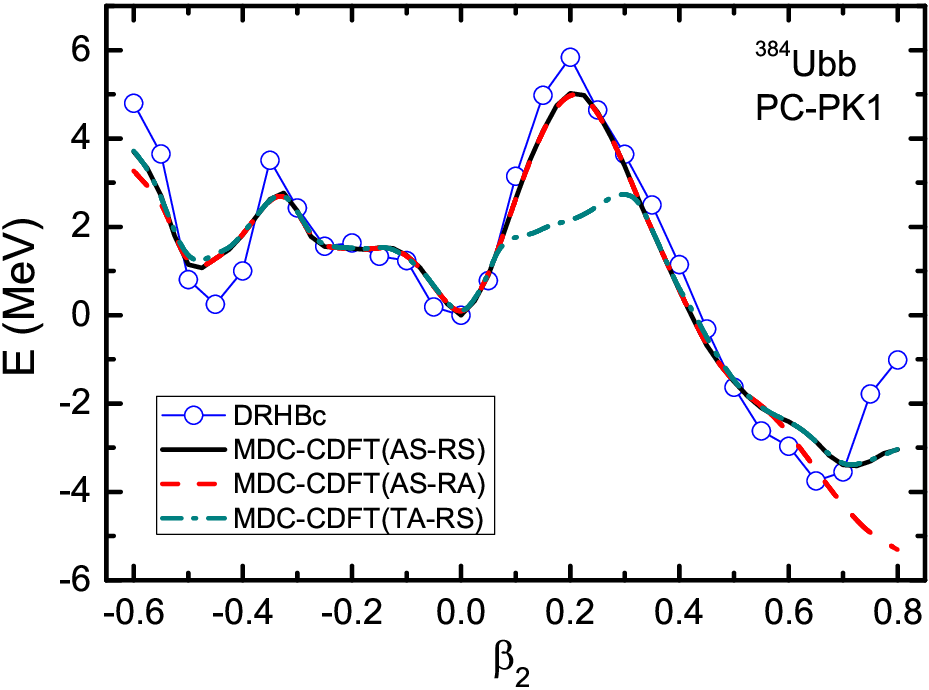}
\caption{The PECs of $^{384}$Ubb obtained by the DRHBc (blue open circles)
and the MDC-CDFT with various self-consistent symmetries imposed:
axially symmetric and reflection symmetric (AS-RS) deformation (black solid curve),
axially symmetric and reflection asymmetric (AS-RA) deformation (red dashed curve)
triaxial and reflection symmetric (TA-RS) deformation (olive dash dotted curve).
The energy at $\beta_2=0$ is subtracted in the DRHBc and MDC-CDFT calculations, respectively.
}
\label{fig:mdc}
\end{figure}

It is well known that for actinides and superheavy nuclei, the inner fission barrier is
generally lowered when triaxial deformation is allowed,
while the outer barrier is further reduced when octupole deformation
is considered~\cite{Damgaard1969_NPA135-432, Moller1970_PLB31-283,
Girod1983_PRC27-2317, Rutz1995_NPA590-680, Abusara2010_PRC82-044303, Zhang2024_ChinPhysC48-104105}.
In some cases, the outer barrier may even disappear after including octupole degrees of freedom.
Therefore, it is important to examine whether the minima identified in Fig.~\ref{fig:jm}
remain stable when triaxial and octupole deformations are taken into account.
However, the present DRHBc calculations keep both axial and reflection symmetry.
Although a triaxial version of the DRHBc has recently been
developed~\cite{Zhang2023_PRC108-L041301, Zhang2025_PRC112-044308}, systematic calculations
for superheavy nuclei within this framework remain computationally prohibitive.

The MDC-CDFT developed by Lu et al., can break both axial and
reflection symmetries~\cite{Lu2012_PRC85-011301R, Lu2014_PRC89-014323}.
To examine the effects of octupole and triaxial deformations on the energy minima of $Z=122$ isotopes,
we still take $^{384}$Ubb as an example and calculate its PECs
using MDC-CDFT under different self-consistent symmetry constraints, as shown in Fig.~\ref{fig:mdc}.
The black solid curve represents the result with axially symmetric and reflection-symmetric (AS-RS) deformation,
the red dashed curve corresponds to the axially symmetric but reflection-asymmetric (AS-RA) case,
and the olive dash-dotted curve shows the result with triaxial and reflection-symmetric (TA-RS) deformation.
Results obtained from the DRHBc calculations are also plotted as blue open circles for comparison.
For clarity, the energy at $\beta_{2}=0$ has been subtracted in both the DRHBc and MDC-CDFT calculations.
It can be seen that the minima corresponding to spherical, oblate, and prolate deformations
are essentially the same between the DRHBc and MDC-CDFT (AS-RS) calculations.
The differences arise from the different pairing interactions and basis choices employed in these two methods.
When reflection symmetry is broken (AS-RA), the prolate minimum near $\beta_{2} \sim 0.7$ disappears.
This indicates that an energy minimum exhibiting very large quadrupole deformation
should not be identified as the ground state, despite possessing the lowest energy.
When axial symmetry is broken (TA-RS), the barrier around $\beta_{2} \sim 0.2$ is lowered.
Nevertheless, these minima still persist, suggesting that the oblate minimum
with $\beta_{2} \sim -0.4$ to $-0.5$ should indeed be considered as the ground state if it possesses the lowest energy.
Note that in the calculated PECs for all the $Z=122$ isotopes,
there is no minimum with even larger oblate deformation ($\beta_{2} <-0.6$).
Whether they can be recognized as ground states when they possess the lowest energy
in other isotopes still needs to be checked carefully.
Following this strategy, the ground states for all the $Z=122$ isotopes
are determined using the unconstrained DRHBc calculations (red solid triangles in Fig.~\ref{fig:pec}) and
the bulk properties, such as binding energies, nucleon separation energies, etc., are obtained.

\begin{figure}[h]
\centering
\includegraphics[width=0.95\columnwidth]{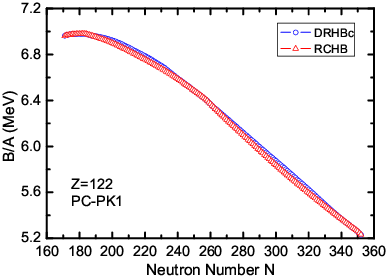}
\caption{The binding energy per nucleon $B/A$ for
the ground states of $Z = 122$ isotopes as a function of neutron
number calculated by the DRHBc (blue solid line) and RCHB (red dashed line)
using the density functional PC-PK1.
}
\label{fig:ba}
\end{figure}

\begin{figure}[h]
\centering
\includegraphics[width=0.95\columnwidth]{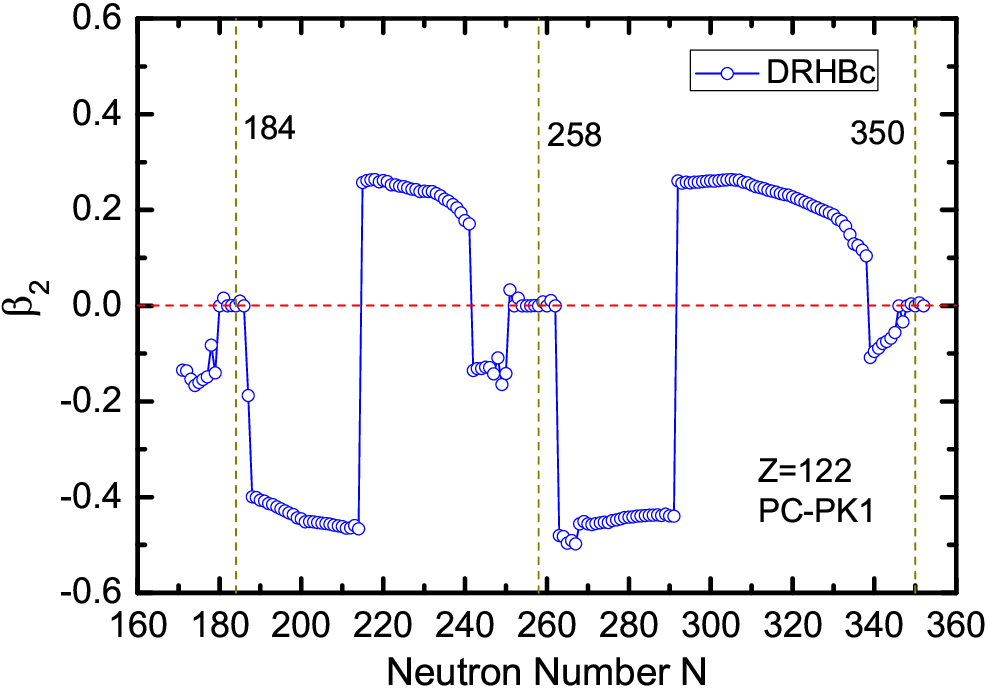}
\caption{The quadrupole deformation $\beta_2$ for the ground states of $Z = 122$
isotopes as a function of neutron number calculated by the DRHBc (blue solid circles)
using the density functional PC-PK1.
}
\label{fig:de}
\end{figure}

Figure~\ref{fig:ba} shows the binding energy per nucleon $B/A$
for the ground states of $Z = 122$ isotopes as a function of neutron number,
calculated within both the DRHBc (blue solid line)
and the spherical RCHB (red dashed line) frameworks.
A local maximum around $N \approx 180$ is observed in the results of both models.
This feature is similar to that found for isotopes of $Z=117$-120~\cite{Zhang2024_PRC110-024302},
a behavior primarily governed by the competition between the volume energy
contribution and the contributions from the Coulomb and asymmetry energies.
The DRHBc and RCHB results are in close agreement in the neutron-number
regions $N = 171$-186, 242-262, and 339-352.
This consistency stems from the spherical or weakly oblate deformed shapes
of the nuclei in these regions, as shown in Fig.~\ref{fig:de}.
When deformation is included self-consistently in the DRHBc calculations,
the nuclei generally gain additional binding energy compared to the spherical RCHB results.

\begin{figure}[h]
\centering
\includegraphics[width=0.95\columnwidth]{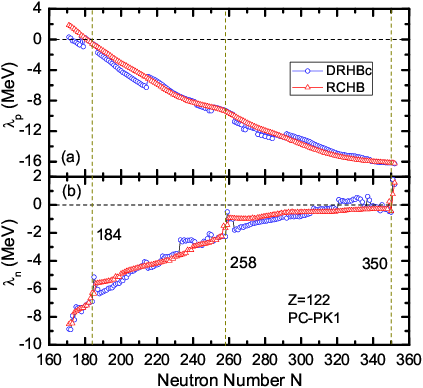}
\caption{The (a) proton Fermi surface $\lambda_p$ and (b) neutron Fermi surface $\lambda_n$
for the ground states of $Z = 122$ isotopes as a function of neutron
number calculated by the DRHBc (blue open circles) and the RCHB (red open triangles)
using the density functional PC-PK1.
}
\label{fig:fe}
\end{figure}

\begin{figure}[h]
\centering
\includegraphics[width=0.95\columnwidth]{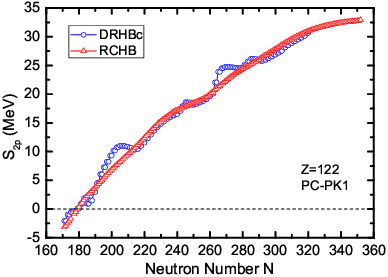}
\caption{The two-proton separation energy $S_{2p}$ for the ground states of $Z = 122$
isotopes as a function of neutron number calculated by the DRHBc (blue open circles)
and the RCHB (red open triangles) using the density functional PC-PK1.
}
\label{fig:sp}
\end{figure}

Figure~\ref{fig:fe} shows the proton ($\lambda_p$) and neutron ($\lambda_n$) Fermi energies
for the ground states of $Z = 122$ isotopes as a function of neutron number calculated by the DRHBc.
Results from the spherical RCHB theory are also shown for comparison.
As shown in Fig.~\ref{fig:fe}(a), the proton Fermi energy $\lambda_p$ obtained
in the RCHB calculation decreases smoothly with neutron number and becomes negative beyond $N=181$.
We know that the $\lambda_p$ for the nuclei within the proton drip-line should be negative.
However, whether the nucleus is bound also depends on the nucleon separation energies,
which are positive for a bound nucleus.
From Fig.~\ref{fig:sp}, the two-proton separation energy $S_{2p}$
calculated within RCHB is negative for $N \le 178$.
Combining these criteria, the RCHB calculation indicates that $^{303}$Ubb ($N=181$)
is the proton drip-line nucleus, with isotopes having $N \le 180$ being unbound.
In the DRHBc results, $\lambda_p$ shows an overall decreasing trend with neutron number
but exhibits abrupt increases at specific neutron numbers.
Notably, $\lambda_p$ first turns negative at $N \ge 173$,
then becomes positive again at $N=180$ and 181.
The two-proton separation energy $S_{2p}$ from DRHBc (Fig.~\ref{fig:sp}) increases with neutron number,
albeit with small fluctuations, and is negative for $N \le 178$, consistent with the RCHB results.
Therefore, according to the DRHBc calculations, isotopes with $N \le 181$ are unbound,
and the proton drip-line nucleus is $^{304}$Ubb ($N=182$).

\begin{figure}[h]
\centering
\includegraphics[width=0.95\columnwidth]{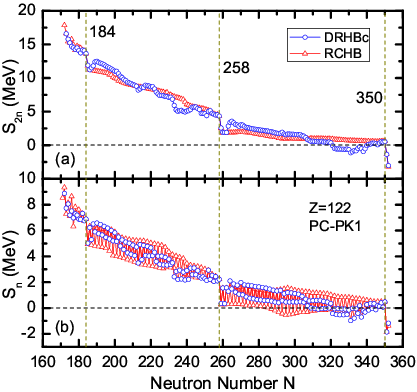}
\caption{The same as Fig.~\ref{fig:sp}, but for (a) the two-neutron separation energy $S_{2n}$
and (b) the one-neutron separation energy $S_{n}$.
}
\label{fig:sn}
\end{figure}

As for the neutron Fermi energy $\lambda_n$ [see Fig.~\ref{fig:fe}(b)], it continuously increases
with neutron number both in the RCHB and DRHBc calculations, exhibiting sharp rises at
specific neutron numbers such as $N=184$, 258, and 350.
These abrupt increases suggest the existence of neutron shell closures.
The present calculations identify $N=184$, 258, and 350 as possible neutron magic numbers,
consistent with previous DRHBc studies~\cite{Du2024_Particles7-1086, Liu2024_Particles7-1078,
Pan2025_Particles8-2, Wu2025_Particles8-19}.
It can be seen that in the DRHBc results, $\lambda_n$ becomes positive for $N=321$-334,
turns negative again until $N=351$, but remains positive for a few nuclei at $N=337$, 338, and 345.
Within the spherical RCHB framework, this  phenomenon is less pronounced.
The $\lambda_n$ only becomes positive at $N=349$ before the shell closure $N=350$.
From Fig.~\ref{fig:sn}(a) it can be seen that the two-neutron separation energy $S_{2n}$
calculated by the DRHBc generally decreases with neutron number (with minor fluctuations)
and drops sharply precisely at $N=184$, 258, and 350,
which are coincide with those sudden increases in the neutron Fermi energy.
This correlation confirms $N=184$, 258, and 350 as neutron magic numbers.
The two-neutron separation energy $S_{2n}$ values become negative for $N=321$-340
and positive again from $N=341$ to the shell closure $N=350$,
indicating that nuclei with $N=321$-340 are unbound.
This identifies the neutron drip-line for $Z=122$ isotopes in the DRHBc as $^{442}$Ubb ($N=320$).
However, several nuclei far away from the neutron drip-line may still exist.
In contrast, the RCHB predicts $S_{2n}$ to remain positive until beyond the shell closure $N=350$,
suggesting more bound nuclei than the DRHBc.
This indicates that treating the deformation effects as well as pairing and
continuum effects in a consistent way can affect significantly the predicted location of
neutron drip-line~\cite{In2021_IJMPE30-2150009}.
The one-neutron separation energy $S_n$ should also be positive if the nucleus is bound.
In Fig.~\ref{fig:sn}(b), it can be seen that $S_n$ also decreases with neutron number
and shows sharp drops at the same magic numbers.
A clear odd-even staggering is observed in both DRHBc and RCHB calculations.
In the DRHBc, the $S_n$ for the nuclei at $N=$311, 313, 315, 317, 341, 343, 345 are negative.
This illustrates that several nuclei far away from neutron drip-line $N=320$ are bound
since their Fermi energies are negative, and their one- and two-neutron separation energies are positive.
This phenomenon is quite similar with the so called ``stability peninsula'' identified in the
$50\le Z \le70$~\cite{Pan2021_PRC104-024331} and
$100\le Z \le120$~\cite{Zhang2021_PRC104-L021301, He2021_ChinPhysC45-101001, He2024_PRC110-014301} regions.
In the RCHB results, $S_n$ is negative for odd-$N$ nuclei in the region $N=287$-341,
marking them as unbound.
According to the $S_n$ and $S_{2n}$ systematics in the RCHB,
the neutron drip-line is located at $^{472}$Ubb ($N=350$).

\begin{figure}[h]
\centering
\includegraphics[width=0.95\columnwidth]{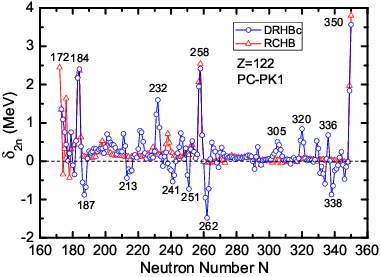}
\caption{Two-neutron shell gaps $\delta_{2n}$ for the ground states of $Z = 122$
isotopes as a function of neutron number calculated by the DRHBc (blue open circles)
and the RCHB (red open triangles) using the density functional PC-PK1.
}
\label{fig:del2n}
\end{figure}

To better illustrate the shell closures and the fine structure of $S_{2n}$ in the $Z = 122$ isotopes,
Fig.~\ref{fig:del2n} presents the two-neutron shell gaps $\delta_{2n}$ as a function of neutron number,
calculated using the DRHBc and RCHB models.
$\delta_{2n}$ is defined as the difference in $S_{2n}$ between two neighboring nuclei:
\begin{equation}
\delta_{2n} (Z, N) = S_{2n}(Z, N)- S_{2n}(Z, N+2).
\end{equation}
The overall patterns from the DRHBc and RCHB results are quite similar.
At the known or possible shell closures $N = 184$, $258$, and $350$,
both models yield pronounced peaks in $\delta_{2n}$.
The DRHBc calculations exhibit more small peaks than the RCHB ones,
which can be attributed to deformation effects omitted in the RCHB approach.
Most $\delta_{2n}$ values from both models are non-negative,
reflecting the globally decreasing trend of $S_{2n}$ with increasing neutron number.
The DRHBc results show more negative peaks than the RCHB ones.
The only two small negative peaks in the RCHB results correspond to nuclei beyond the proton drip line,
whereas the negative peaks in the DRHBc results are associated with the fine structure of $S_{2n}$,
indicating shape changes in these isotopes (see Fig.~\ref{fig:de}).
Therefore, it is necessary to examine the evolution of quadrupole deformation in the $Z = 122$ isotopes.

Figure~\ref{fig:de} shows the quadrupole deformation $\beta_2$ for the ground states
of $Z = 122$ isotopes as a function of neutron number calculated by the DRHBc.
Similar evolutionary patterns can be observed between the predicted magic numbers,
specifically in the regions from $N=184$ to $258$ and from $N=258$ to $350$.
First, the ground states are spherical or nearly spherical near the magic numbers.
Then, with the addition of a few neutrons, they rapidly develop a highly
oblate deformation ($\beta_{2} \sim -0.4$ to $-0.5$ in the ranges $N=188$-$214$ and $N=263$-$291$).
Subsequently, the nuclei shift to a moderately prolate deformation ($\beta_{2} \sim 0.3$),
which then decreases to $\beta_{2} \sim 0.1$ as the neutron number increases further.
This is followed by a transition to a weakly oblate shape ($\beta_{2} \sim -0.1$).
Finally, the spherical shape is restored upon approaching the next magic number.
It can also be seen that shape transitions occur at $N = 187$, $215$, $242$,
$251$, $263$, $292$, and $339$, most of which correspond to or lie near the
small peaks in the two-neutron shell gap $\delta_{2n}$ shown in Fig.~\ref{fig:del2n}.

\begin{figure}[h]
\centering
\includegraphics[width=0.95\columnwidth]{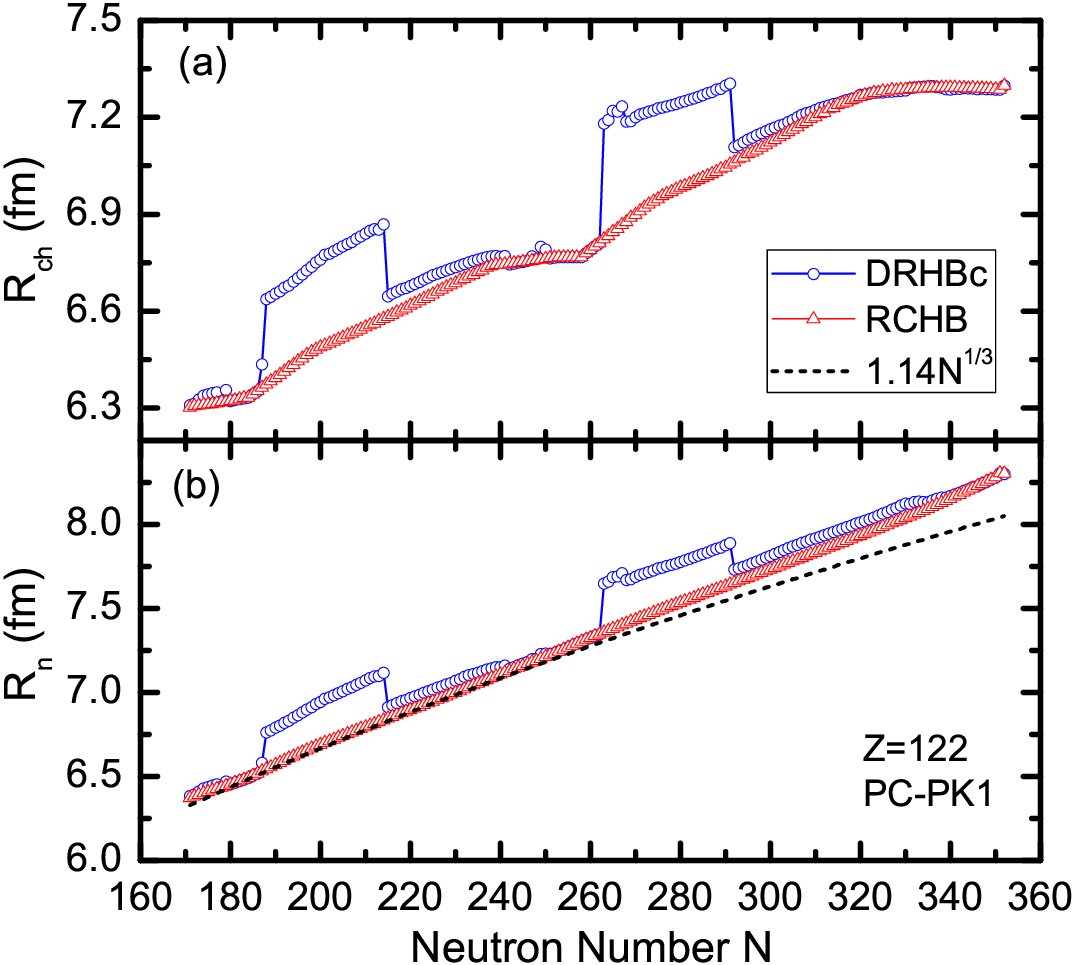}
\caption{(a) The charge radii $R_{\rm ch}$
and (b) the neutron radii $R_n$ for
the ground states of $Z = 122$ isotopes as a function of neutron
number calculated by the DRHBc (blue open circles) and the RCHB (red open triangles)
using the density functional PC-PK1. The results obtained by the empirical formula
$R_n=1.14N^{1/3}$ is shown as dashed line.
}
\label{fig:ch}
\end{figure}

Figure~\ref{fig:ch} shows the charge radii ($R_{\rm ch}$) and neutron radii ($R_n$)
for the ground states of $Z = 122$ isotopes as a function of neutron number,
calculated by the DRHBc (blue open circles) and the RCHB (red open triangles) theories.
For comparison, results from the empirical formula $R_n = r_0N^{1/3}$ with $r_0=1.14$ determined by
the neutron rms radius of $^{208}$Pb~\cite{Guo2024_ADNDT158-101661} are also shown (dashed line).
The empirical formula reproduces the RCHB neutron radii well at smaller neutron numbers,
but the deviation grows larger with increasing neutron number.
This suggests that isospin-dependent effects become important
for accurately describing neutron radii in this region.
The $R_{\rm ch}$ and $R_n$ values obtained from the DRHBc and RCHB calculations
agree well near the neutron magic numbers $N = 184$, 258, and 350,
where the nuclei are spherical in the DRHBc picture.
For nuclei with small deformations ($|\beta_{2}| < 0.3$),
the results from the two methods are also quite close.
Nearly all DRHBc results are larger than the corresponding RCHB values,
except for a few nuclei around $N \sim 340$ with oblate deformation.
A similar trend has been observed in earlier DRHBc studies and is
attributed to specific shell structures, particularly key single-particle levels
near the Fermi surface~\cite{Pan2025_PRC112-024316}.
For nuclei with large oblate deformation,
the DRHBc results are significantly larger than those from the RCHB.

\begin{figure*}[!]
\centering
\includegraphics[width=0.7\textwidth]{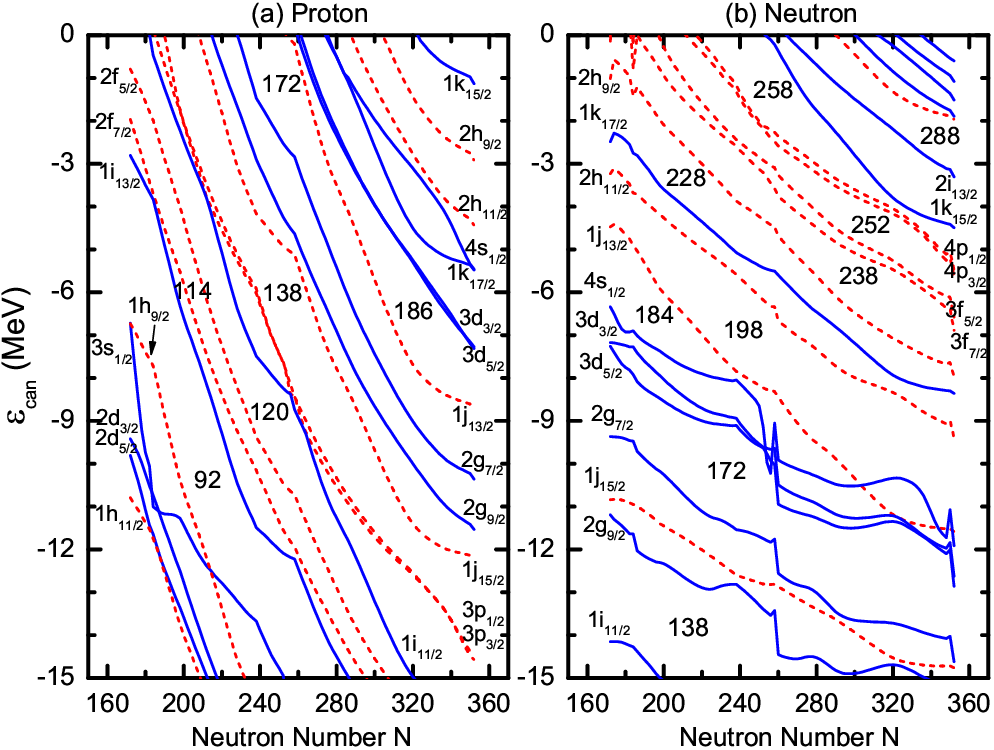}
\caption{Single-particle energy levels for spherical states of $Z=122$ isotopes,
shown separately for (a) protons and (b) neutrons in the canonical basis,
as a function of neutron number. The positive and negative parity levels
are denoted by blue solid and red dashed lines, respectively.
}
\label{fig:spl}
\end{figure*}

Figure~\ref{fig:spl} displays the single-particle energy levels for spherical configurations
of $Z=122$ isotopes in the canonical basis as a function of neutron number.
The proton levels exhibit relatively smooth evolution with neutron number,
except for a limited number of discrete states.
Notably, the $Z=120$ shell gap remains nearly constant across neutron numbers.
Due to the intruder orbital $1i_{11/2}$, this gap becomes largest around $N=258$,
which predicts as a neutron magic number.
Additionally, significant spherical shell gaps are observed at $Z=92$ and $Z=138$ in Fig.~\ref{fig:spl}(a).
The $Z=92$ shell gap contradicts with experimental data and corresponds to a spurious
shell closure in certain relativistic mean-field parametrizations~\cite{Rutz1998_NPA634-67},
though this artifact can be mitigated by density functional updates~\cite{Long2007_PRC76-034314}.
The $Z=138$ shell gap arises from splitting in the proton $3p$ states,
where reduced splitting correlates with enhanced gap magnitude.
Moreover, there is a relatively smaller spherical
shell gap at $Z = 114$ formed by the splitting of proton $2f$ states,
and no discernible gap appears at $Z=126$ in any isotope within the present DRHBc calculations.

For neutron levels in Fig.~\ref{fig:spl}(b), abrupt energy shifts occur for specific orbitals
at neutron shell closures $N=184$, 258, and 350, especially for the positive parity levels.
Multiple spherical shell gaps can be identified.
The $N=184$ shell gap diminishes with increasing neutron number owing to
the $4s_{1/2}$ state and the intruder state $1j_{13/2}$, whereas the $N=172$ shell gap remains robust.
It weakens significantly only around neutron number $N=258$.
It can be seen that the magnitude of the $N=172$ shell gap exhibits an inverse correlation with
the spin-orbit splitting in $2f$ and $3d$ states: reduced splittings correspond to enhanced gap stability.
Similarly, the $N=258$ shell gap decreases with neutron number due to the $1k_{15/2}$ intruder orbital.
It nearly vanishes around the neutron drip line $N=320$.
Note that the $N=350$ shell gap is not displayed as it resides in the positive energy continuum,
where resonant states dominate.

\begin{figure}[h]
\centering
\includegraphics[width=0.95\columnwidth]{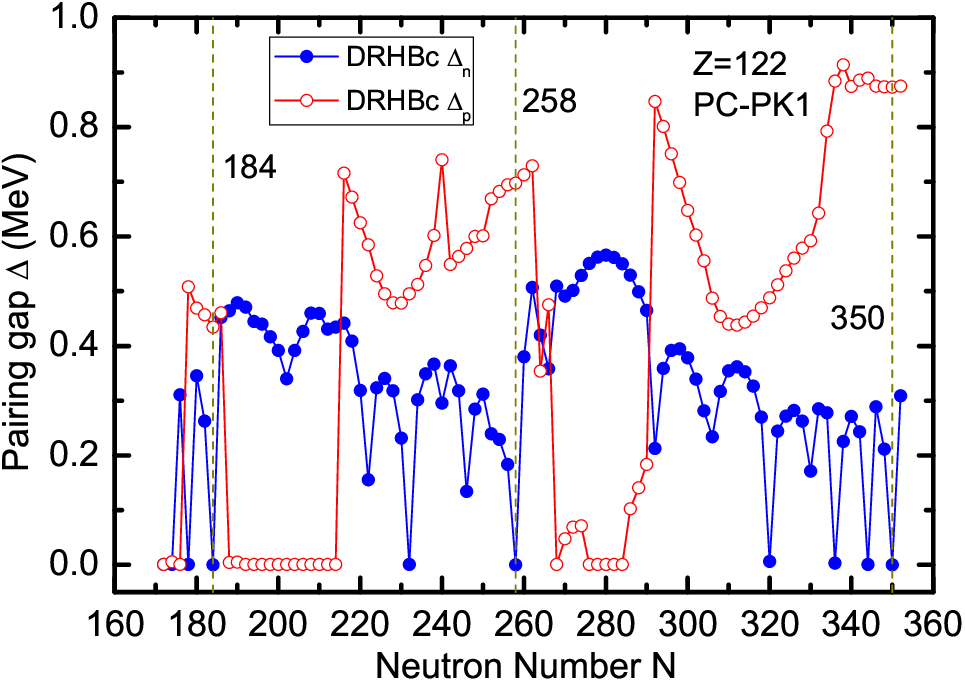}
\caption{The proton and neutron pairing gap, $\Delta_p$ (red open circles) and $\Delta_n$ (blue solid circles),
for the ground states of even-even $Z = 122$ isotopes as a function of neutron
number calculated by DRHBc using the density functional PC-PK1.
}
\label{fig:delta}
\end{figure}

The average pairing gap $\Delta$ provides detailed information about the impact of pairing correlations
and serves as an important indicator of possible shell closure.
Fig.~\ref{fig:delta} shows the proton pairing gap $\Delta_p$ and neutron pairing gap $\Delta_n$
for the ground states of even-even $Z = 122$ isotopes as a function of neutron number, as calculated by DRHBc.
The neutron pairing gap $\Delta_n$ drops to zero at the candidate shell closures $N = 184$, $258$, and $350$,
as well as at specific neutron numbers such as $N = 232$, $320$, $336$, and $344$, which may correspond to subshell closures.
In addition, $\Delta_n$ also shows decreases (not to zero) at several positions.
Note that these subshell structures are not discernible in Fig.~\ref{fig:spl}(b)
due to either their origin in deformed shell gaps or their location within the positive energy region.
Between the shell and subshell closures, $\Delta_n$ first increases and then decreases, exhibiting arch-like structures.
The average pairing gap also reflects nuclear deformation, as the mean field is influenced by pairing correlations.
Since $Z = 120$ is a magic number [see Fig.~\ref{fig:spl}(a)] predicted by the covariant density functional theory~\cite{Bender2003_RMP75-121},
proton pairing collapse may still occur when two more protons are added for the $Z = 122$ isotopes.
The proton pairing gap $\Delta_p$ becomes zero or very small in several mass regions.
Interestingly, these regions coincide exactly with those where the nuclei exhibit large oblate deformation (see Fig.~\ref{fig:de}).
In regions where $\Delta_p$ is relatively larger, it first decreases and then increases,
exhibiting inverted arch-like structures that contrast with the behavior of $\Delta_n$.
As can be seen in Fig.~\ref{fig:de}, the nuclei in these regions are all prolate deformed, weakly oblate deformed, or spherical.

\section{Summary}\label{sec:summary}

In this work, the ground-state properties of superheavy $Z=122$ isotopes are
investigated using the deformed relativistic Hartree-Bogoliubov theory in continuum.
Bulk properties of the ground states, including binding energies,
Fermi energies, nucleon separation energies,
quadrupole deformations $\beta_2$, two-neutron shell gaps $\delta_{2n}$,
root-mean-square radii, and average pairing gaps are calculated.
The results are also compared with those obtained from
the spherical relativistic continuum Hartree-Bogoliubov theory.
By examining the effects of angular momentum cutoff $J_{\text{max}}$, triaxial deformation,
and octupole deformation, a strategy for determining the ground states is proposed.
The present investigation shows that an angular momentum cutoff of $J_{\text{max}} = 31/2~\hbar$
is sufficient for the $Z=122$ isotopes.
The inclusion of octupole deformation eliminates the large prolate minimum,
while triaxial deformation only lowers the first barrier on the prolate side,
leaving the minima in the potential energy curve unchanged.
Furthermore, based on an analysis of Fermi and separation energies,
the proton and neutron drip-lines for $Z = 122$ isotopes are determined
within both the DRHBc and RCHB frameworks.
Possible neutron magic numbers at $N=184$, $258$, and $350$ are also suggested.
Finally, the evolution of the two-neutron shell gaps, deformation, charge and neutron radii,
single-particle levels, and average pairing gaps with increasing neutron number is discussed.
These quantities consistently support the suggested neutron shell closure.

\begin{acknowledgements}

Helpful discussions with members of the DRHBc Mass Table Collaboration are gratefully acknowledged.
This work is supported by the National Natural Science Foundation of China (Nos. 11875027, 12475121, 12205097, 12305125),
the National Key Laboratory of Neutron Science and Technology (Grant No. NST202401016),
and the High-Performance Computing Platform of North China Electric Power University.

\end{acknowledgements}


%

\end{document}